\documentstyle[amssymb,eqsecnum,aps,epsf]{revtex}

\begin{document}
\draft
\title{Spatial string tension in gluodynamics at finite temperature{\tt \ }in a
spherical model approximation}
\author{V.~K.~Petrov\thanks{
E-mail address: petrov@earthling.net}}
\address{Bogolyubov Institute for Theoretical Physics,\\
National Academy of Sciences of Ukraine, Kiev 143, UKRAINE}
\maketitle

\begin{abstract}
Spatial string tension is computed in finite temperature gluodynamics\ on
asymmetric lattices in a spherical model approximation. Conditions of
scaling behavior are specified. Discrepancies with a standard
renormalisation procedure are discussed.
\end{abstract}


\narrowtext

\section{Introduction}

Recently \cite{bali1,bali2,borgs,karsch,caselle,teper} the behavior of the
spatial string tension was studied in $(d+1)$-dimensional ($d=2,3$ ) $SU(2)$
and $SU(3)$ gauge theories. The spatial string tension $\sigma $ at high
temperature in $(3+1)$-dimensional $SU(N)$ gauge theory was rigorously
proved \cite{borgs} to be non-vanishing at finite lattice spacing. The
spatial string tension, as it was pointed out by \cite{svetitsky}, was not
related to the confining properties of a physical potential in the $(3+1)$%
-dimensional theory. The reason is that under $Z\left( N\right) $\ -
transformation topologically trivial Wilsons loops remain invariant; on the
contrary, topologically non-trivial loops such as Polyakov lines transform
as $\Omega \left( {\bf x}\right) \rightarrow z\Omega \left( {\bf x}\right) .$%
\ Therefore the behavior of topologically trivial Wilsons loops cannot be
considered as a confinement criterion \cite{svetitsky}. In particular, the
expectations of a large space-like Wilsons loop may show area law behavior
without static quarks being confined.

Despite the fact that the study of space-like Wilsons loops behavior at
finite temperatures\ does not give straightforward information about
critical phenomena in LGT, it helps to understand better non-perturbative
effects that manifest themselves in correlation functions for the spatial
components of gauge fields.

The remarkable feature of the spatial string tension\ is that it is\ scaling
of $\frac{\sqrt{\sigma }}{T_c}$ \cite{karsch} and thus $\frac{\sqrt{\sigma }%
}{T_c}$\ is non-vanishing in the continuum limit. The calculations of an
average value of a time- and space-like Wilsons loop at $g^2\sim
g_{critical}^2$ i.e. at fixed cut-off have been performed for$\ $a broad
temperature interval ($T$ was varied by varying $N_\tau $) for $SU\left(
2\right) $ \cite{bali2} and for $SU\left( 3\right) $ \cite{karsch} gauge
groups. It was shown that the spatial string tension remained temperature
independent up to $T_c$ and than was rising rapidly, unlike the temporal one
that decreased with temperature above $T_c$. Similar behavior has been found
in lower dimensions and also in $Z(2)$ gauge theory \cite
{bali1,caselle,teper,kar}.

The main features of high temperature behavior of such observables as the
heavy quark potential and spatial string tension can be understood in terms
of the structure of the effective, three-dimensional theory which was
obtained from dimensional reduction at high temperature by means of
perturbation theory \cite{reisz}.\ The basic suggestion is that at high
temperatures temporal dimension becomes arbitrary small and degrees of
freedom in that direction are frozen, therefore $\sum_{t=0}^{N_\tau }\sum_{%
{\bf x}}S\approx \frac 1{a_\tau \cdot T}\sum_{{\bf x}}S$\ so for three
dimensional couplings one can write 
\begin{equation}
\beta _{(3)}\equiv N_\tau \beta =\frac 1{a_\tau T}\frac{2N}{g^2\left(
T\right) }=\frac 1{a_\tau }\frac{2N}{g_{\left( 3\right) }^2},  \label{g3-g4}
\end{equation}
with 
\begin{equation}
g_{(3)}^2=g^2(T)T.
\end{equation}

As it was established in \cite{karsch}, the Higgs part of such an effective
theory does not contribute substantially in the spatial string tension,
leading to the simple relation between $\sigma _3$ and $\sigma $. One of the
remarkable results of \cite{karsch} can be given as 
\begin{equation}
\sqrt{\sigma _3}\approx cN\cdot g_{(3)}^2;~~~~~c\sim \frac 15;~~~N=2;3.
\end{equation}
It means that $\sqrt{\sigma _3}/T_c\approx cN\frac{g_{\left( 3\right) crit}^2%
}{T_c}=cNg_{crit}^2$\ is in agreement with the MC experiment, which shows
that the scaling violation of the ratio $\sqrt{\sigma }/T_c$\ is small
enough \cite{bali1}.

Though the results obtained in a MC simulation have already answered many
crucial questions, we hope that an attempt of an analytic study presented
here will be useful for a more detailed understanding of a scaling
phenomenon. The present paper is organized as follows. In Sect.2 we discuss
the main suggestions, that have been made in the model and compute the
average value of a spatial Wilson loop in given approximations. In Sect.3 we
discuss the result, obtained for spatial string tension. In Section 4 we
make an attempt to give a more comprehensive discussion of the spherical
model approximation accuracy and list some standard and nonstandard examples
of model applications.

\section{Model}

To compute an average value of the space Wilson loop $W_{R\times L}(\beta
_\sigma )=\left\langle {\rm W}_{R\times L}\right\rangle $, we shall use the
asymmetrical lattice $\frac{a_\sigma }{a_\tau }=\xi \neq 1$ so for the
action one can write 
\begin{eqnarray}
S &=&S_E+S_H;~~~  \nonumber \\
S_H &=&-\beta _\sigma \sum_{t=0}^{N_\tau -1}\sum_{\vec x,t,m\neq n}\frac 1N%
{\rm ReSp}\Box _{mn} \\
S_E &=&-\beta _\tau \sum_{t=0}^{N_\tau -1}\sum_{{\bf x},n}\frac 1N{\rm ReSp}%
\Box _{0n};  \label{S-t}
\end{eqnarray}
with 
\begin{equation}
\beta _\sigma =\beta _{nm}=\frac{2N}{g_\sigma ^2\xi };~~\beta _\tau =\beta
_{0n}=\frac{2N\xi }{g_\tau ^2}.  \label{asymm}
\end{equation}
and 
\begin{equation}
\Box _{\mu \nu }=U_\mu \left( x\right) U_\nu \left( x+\mu \right) U_\mu
^{\dagger }\left( x+\nu \right) U_\nu ^{\dagger }\left( x\right) 
\end{equation}
Lattice spaces $a_0$ and $a_\sigma $ can be made arbitrary small at any
fixed $\xi $ and no special assumptions are made about $g_\tau ^2$ and $%
g_\sigma ^2$ except that their values are wholly determined by certain
renormalisation group equations at any given $a_\tau $, $a_\sigma $, $N_\tau
,$ and with $N_\sigma ^3\equiv N_1N_2N_3.$ With decreasing $\xi $ the
electric part $(S_E)$ of the action becomes negligibly small in comparison
with the magnetic one $(S_H)$, then in $\xi <<1$ limit (which in a way is
opposite to the Hamiltonian limit: $\xi >>1$) it may be ignored. The
magnetic part of the action in this case is split into a set of independent
time slices. In other words, the temporal degrees of freedom are frozen in
the limit $\xi <<1$ (which is natural at high temperatures \cite{reisz}),
and we have 
\begin{eqnarray}
-S_H &=&\beta _\sigma \sum_{t=0}^{N_\tau -1}\sum_{\vec x,t,m\neq n}\frac 1N%
{\rm ReSp}\Box _{mn}  \nonumber \\
&\simeq &\beta _\sigma \cdot N_\tau \sum_{\vec x,m\neq n}\frac 1N{\rm ReSp}%
\Box _{mn}.  \label{sb-n}
\end{eqnarray}
The Wilson loop is placed in one of the slices $t=t_0$ and is not affected
by any other slice, therefore this specific part $(t=t_0)$ of the action
works as an effective action while calculating the average value of Wilson
loop. To put it differently 
\begin{equation}
Z=\prod_t{\rm Sp}_t\left( \exp \left\{ \frac 2{g_\sigma ^2\xi }\sum_{{\bf x}%
,m\neq n}{\rm ReSp}\Box _{mn}\right\} \right) =Z\left( t_0\right) ^{N_\tau },
\end{equation}
and $\beta _\sigma =\frac{2N}{g_\sigma ^2\xi }$ can be regarded as the
effective coupling. Therefore 
\begin{eqnarray}
\  &&W_{R\times L}(\beta _\sigma )=  \label{Wilson} \\
&&\ \ \frac 1{Z\left( t_0\right) }{\rm Sp}_{t_0}\left( {\rm W}_{R\times
L}\exp \left\{ \frac 2{g_\sigma ^2\xi }\sum_{\vec x,m\neq n}{\rm ReSp}\Box
_{mn}\right\} \right) .  \nonumber
\end{eqnarray}

\subsection{Approximation $SU(N)\simeq Z(N)$}

It is commonly believed that the $Z(N)$ degrees of freedom are responsible
for many important aspects of $SU(N)$ gluodynamics phase structure. The
lattice gauge theory in the vicinity of phase transition is widely known to
show large degree of universality\cite{svetitsky}. The main evidence in
favor of universality is given by the Wilson loop behavior, whose functional
form does not depend on the choice of a gauge group or on the specifics of
ultraviolet behavior of the model, showing rather simple dependance on space
dimension. Universality arguments place the finite - temperature $SU(N)$ -
gauge theory in the universality class of globally $Z(N)$ invariant systems
with short - range interactions. Hence it is convenient to study such
universal infrared behavior in $SU\left( N\right) \backsim Z\left( N\right) $
approximation: 
\begin{equation}
U_{x;\mu }\simeq z_{x;\mu }=\exp \left\{ \frac{2\pi iq_{x;\mu }}N\right\}
~;~~q_{x;\mu }=0,\dots ,N-1.  \label{z-appr}
\end{equation}

The MC experiment as well as model calculations \cite{agostini} demonstrate
that $SU\left( 2\right) $ and $Z\left( 2\right) $ spectra show remarkable
agreement not only between the pattern of the states, but also between the
values of the masses (except for the lowest state) (see also \cite{caselle}
and \cite{teper}). Such models with discrete gauge groups are easier to
handle. In particular, the method of duality transformations is just the one
elaborated well enough for the systems with discrete symmetry (\cite{sav}
and references there) These models with $Z(N)$ gauge symmetry are also known
to provide a transparent realization of 't Hooft algebra of order and
disorder operators. Moreover, there is evidence \cite{yon} that the effect
of the quantum fluctuations near to the $Z(N)$ configurations on the
symmetrical lattice leads only to a finite renormalisation of the coupling
constant: 
\begin{equation}
\beta _{old}\rightarrow \beta _{new}=\beta _{old}-\frac{N^2-1}4.  \label{yon}
\end{equation}
Since the additional term in (\ref{yon}) depends neither on $a_{\sigma ,\tau
}$ nor on $\beta $ we may hope that on the asymmetrical lattice the effect
of the quantum fluctuations will also lead again to an insignificant change
of the coupling constant. We consider it to be still another justification
of the chosen approach. Of course, nobody expects full coincidence of the
results for $SU\left( N\right) $ and $Z\left( N\right) .$

One of the main advantages of the $SU\left( N\right) $ $\approx $ $Z\left(
N\right) $ approximation is that one can apply it to $Z\left( N\right) $
duality transformations, which relate $Z\left( N\right) $ gluodynamics to$\ $%
Ising ($N=2$ ) and Potts ($N=3$ ) models in $3$ - dimensional space .

\subsection{Duality transformations}

As it is well known \cite{guth}, in the case of $Z(2)$ gauge group the
Wilson loop $R\times L$ average value placed at $x_1=0$ can be calculated on
the dual lattice 
\begin{equation}
W_{R\times L}=\left\langle \exp \left\{ -2\sum_{R\times L}\beta _n^{\prime
}z_{\vec x}z_{\vec x+n}\right\} \right\rangle ;~~~z_{\vec x}\in Z(2).
\label{Z(2)}
\end{equation}
Summation $\sum_{R\times L}$ is done over all dual Ising spins placed at $%
x_1=0$ inside $R\times L$. In other words, ferromagnetic links dual to
plaquettes of $R\times L$ transform into antiferromagnetic links of the same
strength \cite{guth}. In the $Z(3)$ gauge theory where the coupling $\beta
^{\prime }$ is multiplied by $e^{\frac{2\pi }3i}$ at links dual to
plaquettes placed inside $R\times L$ at $x_1=0$ we have 
\begin{equation}
W_{R\times L}=\left\langle \exp \left\{ -\left( 1-e^{\frac{2\pi }3i}\right) 
{\rm Re}\sum_{R\times L}\beta _n^{\prime }z_{\vec x}z_{\vec x%
+n}^{-1}\right\} \right\rangle .~~~  \label{Z(3)}
\end{equation}

The duality transformations on the asymmetric lattice ($a_1\neq a_2\neq a_3$
and $\beta _1^{\prime }\neq \beta _2^{\prime }\neq \beta _3^{\prime })$
relate dual links $l_n$ to plaquettes $P_{mk}$ of the original lattice $%
(n\neq k\neq m)$ \cite{aver}. The couplings $\beta _n^{\prime }$ on the dual
asymmetric lattice are related to the corresponding ones $(\beta _{km})$ on
the original lattice in the form 
\begin{equation}
\beta _n^{\prime }\approx \left\{ 
\begin{array}{ll}
e^{-\left( 1-\cos \frac{2\pi }N\right) \beta _{km}}; & \beta _{km}>>1, \\ 
\frac 1N\ln \frac 1{\beta _{km}}; & \beta _{km}<<1.
\end{array}
\right.   \label{c-ng}
\end{equation}
In the specific case of $a_n=a$, ($\beta _n^{\prime }=\beta _\sigma ^{\prime
}$ ; $\beta _{km}=\beta _\sigma $ ) 
\begin{equation}
\beta _\sigma ^{\prime }\approx \left\{ 
\begin{array}{ll}
e^{-\left( 1-\cos \frac{2\pi }N\right) \beta _\sigma }; & \beta _\sigma >>1,
\\ 
\frac 1N\ln \frac 1{\beta _\sigma }; & \beta _\sigma <<1.
\end{array}
\right.   \label{c-ng-p}
\end{equation}

\subsection{Spherical model approximation}

To compute the partition function\footnote{%
Sources $\eta _x$ are introduced for convenience.} 
\begin{equation}
Z=\sum_{\left\{ z\right\} }\exp \left\{ {\rm Re}\sum_{x,n}\beta _n^{\prime
}z_xz_{x+n}^{*}+\sum_x\eta _xz_x\right\}   \label{part}
\end{equation}
on the dual lattice we use the well-known spherical model \cite{kac} (see
e.g.\cite{joyce}, \cite{baxter} and references therein). The crucial point
of this model lies in replacing the precise condition $\left| z_x\right| ^2=1
$ by a less burdening ''averaged'' condition 
\begin{equation}
\frac 1{N_\sigma ^3}\sum_x\left| z_x\right| ^2=1.  \label{cond}
\end{equation}
or 
\begin{eqnarray}
\  &&\prod_x\delta (\left| z_x\right| ^2-1)\rightarrow \delta \left(
\sum_xN_\sigma ^3-\left| z_x\right| ^2\right) =  \nonumber \\
&&\ \int_{c-i\infty }^{c+i\infty }\frac{ds}{2\pi i}\exp \left\{ s\left(
\sum_xN_\sigma ^3-\left| z_x\right| ^2\right) \right\} ,  \label{int}
\end{eqnarray}
where the constant $c$ is chosen to the right of all singularities of the
integrand to ensure the correctness of interchanging the integration over $%
z_x$ and $ds$. Now we can change $%
\mathop{\displaystyle \sum }
_{\left[ z_x\right] }\rightarrow \int_{-\infty }^{+\infty }dz_x$ , so the
partition function $\left( \ref{part}\right) $ may be rewritten in the
following form: 
\begin{eqnarray}
\  &&Z\simeq \int \frac{ds}{2\pi i}e^{sN_\sigma ^3}\times   \label{su2-4} \\
&&\ \int_{-\infty }^{+\infty }\prod_xdz_x\exp \left( -\sum_{x,x^{\prime
}}z_xA_{\vec x}^{\vec x^{\prime }}z_{x^{\prime }}+\sum_x\eta _xz_x\right) , 
\nonumber
\end{eqnarray}
where 
\begin{eqnarray}
A_{\vec x}^{\vec x^{\prime }} &=&s\delta _{\vec x}^{\vec x^{\prime
}}-\sum_{n=1}^3\beta _n^{\prime }\delta _{\vec x+n}^{\vec x^{\prime }}\equiv 
\nonumber \\
&&\ m^2\delta _{\vec x+n}^{\vec x^{\prime }}-\sum_{n=1}^3\beta _n^{\prime
}(\delta _{\vec x+n}^{\vec x^{\prime }}-\delta _{\vec x}^{\vec x^{\prime }})
\end{eqnarray}
with 
\begin{equation}
m=\sqrt{s-\sum_{n=1}^3\beta _n^{\prime }}
\end{equation}

The integration over $z_x$ can be now carried out 
\begin{eqnarray}
Z &\simeq &\int ds\exp \left\{ sN_\sigma ^3-\frac 12\ln \det A-\eta _x\left(
A_{\vec x}^{\vec x^{\prime }}\right) ^{-1}\eta _{x^{\prime }}\right\} = 
\nonumber \\
&=&\int ds\exp \left\{ N_\sigma ^3\Phi (s)\right\} .
\end{eqnarray}

To compute the Wilson loop average value, in the partition function $A_{\vec 
x}^{\vec x^{\prime }}=s\delta _{\vec x}^{\vec x^{\prime }}-\sum_{n=1}^3\beta
_n^{\prime }\delta _{\vec x+n}^{\vec x^{\prime }}$ should be changed into $%
A_{\vec x}^{\vec x^{\prime }}+\left( \Delta _{R\times L}\right) _{\vec x}^{%
\vec x^{\prime }}$, where 
\begin{eqnarray}
\left( \Delta _{R\times L}\right) _{\vec x}^{\vec x^{\prime }} &=&2\delta
_n^1\delta _{x_1}^0\delta _{x_1}^{x_1^{\prime }+1}\theta \left( \frac R2%
-x_2\right) \theta \left( x_2+\frac R2\right)   \nonumber \\
&&\cdot \theta \left( x_3+\frac L2\right) \theta \left( \frac L2-x_3\right) .
\label{diff}
\end{eqnarray}
$\cdot $

After the Fourier transformations 
\begin{equation}
\sum_{\vec x}\sum_{\vec x^{\prime }}\exp \left( i\vec \varphi \cdot \vec x-i%
\vec \varphi ^{\prime }\cdot \vec x^{\prime }\right) \cdot A_{\vec x}^{\vec x%
^{\prime }}=A(\varphi _n;\varphi _m^{\prime })
\end{equation}
we have 
\begin{equation}
A(\varphi _n;\varphi _m^{\prime })=N_\sigma ^3\left[ \delta _{\vec \varphi
^{\prime }}^{\vec \varphi }A(\vec \varphi )-\frac{\beta _1^{\prime }}{%
N_\sigma ^3}\Delta (\varphi _n;\varphi _m^{\prime })\cos \varphi _1\right] 
\label{fourier}
\end{equation}
with 
\begin{equation}
A(\vec \varphi )\equiv \sum_{n=1}^3\beta _n^{\prime }\cos \varphi _n,
\label{A}
\end{equation}
and 
\begin{equation}
\Delta (\varphi _n;\varphi _m^{\prime })=2\frac{\sin \frac{(R+1)(\varphi
_2-\varphi _2^{\prime })}2}{\sin \frac{\varphi _2-\varphi _2^{\prime }}2}%
\frac{\sin \frac{(L+1)(\varphi _3-\varphi _3^{\prime })}2}{\sin \frac{%
\varphi _3-\varphi _3^{\prime }}2}.  \label{delta}
\end{equation}

The integration over $s$ can be done by the steepest descent method, the
saddle point $s_0$ being defined by the condition $\left[ \frac \partial {%
\partial s}\Phi (s)\right] _{s=s_0}=0$ which can be rewritten as 
\begin{equation}
1=\frac 12\int_0^{2\pi }\frac{d^3\varphi }{(2\pi )^3}\frac 1{%
s_0-\sum_{n=1}^3\beta _n^{\prime }\cos \varphi _n}.  \label{saddle}
\end{equation}
In the symmetric case the equation $(\ref{saddle})$ has a simple solution in
the critical point vicinity 
\begin{equation}
s_0\approx 3\beta ^{\prime }+2\pi ^2\beta ^{\prime }(\beta _c^{\prime
}-\beta ^{\prime })^2  \label{S0}
\end{equation}
with 
\[
\beta ^{\prime }=\frac 13\sum_{n=1}^3\beta _n^{\prime }
\]
or 
\begin{equation}
m=\sqrt{s_0-3\beta ^{\prime }}\approx \pi \sqrt{2\beta ^{\prime }}(\beta
_c^{\prime }-\beta ^{\prime })\theta (\beta _c^{\prime }-\beta ^{\prime }).
\label{mass}
\end{equation}

The number of sites spanned by the Wilson loop is $1/N_2N_3$ times smaller
than the whole volume, so their contribution doesn't influence the saddle
point position.

To clarify the engine of spherical model approach we would note that $\left( 
\ref{cond}\right) $fixing the compactness condition $\left| z\right| =1$
with shown approximation\ brings(through $\left( \ref{int}\right) $\ ) an
additional 'mass' term into action: 
\begin{equation}
m_0^2=-\sum_{n=1}^3\beta _n^{\prime }\rightarrow m^2=s_0(\beta _n^{\prime
})-\sum_{n=1}^3\beta _n^{\prime }.
\end{equation}
The effective mass (\ref{mass}) is defined by the saddle point condition (%
\ref{saddle}) and plays the role of the screening mass (expressed in lattice
units) in the correlation functions.

As it is well known, theories with the local gauge symmetry are described in
terms of nonlocal order parameter. Thus the partition function also may have
no singularities in the thermodynamic limit, the quantities which determine
the nonlocal order parameters may have singularities as a result of
increasing their size to infinity. This is strongly suspected to occur in
lattice gauge theories for Wilson-loop order parameters, and poses an
obstacle to the strong-coupling expansion \cite{kowall}. Therefore, it seems
quite useful to study differences in behavior of a 'tiny' (one-plaquettes)
and 'large' (whole $N_{{\bf 2}}\times N_{{\bf 3}}$ plain) Wilson loops. In
particular for the 'small' one $\left( 1<<R<<N_{{\bf 2}};\quad 1<<L<<N_{{\bf %
3}}\right) $ it would be enough to take into account the term $\frac{\Delta
\left( \varphi _n;\varphi _m^{\prime }\right) }{N_\sigma ^3}\cos \varphi _1$
in the first order 
\begin{eqnarray}
\  &&\frac{Z_W}{N_\sigma ^3}=\frac{Z_0}{N_\sigma ^3}W_{R\times L}\approx -%
\frac 12\sum_{\vec \varphi }\ln \left[ s_0-A(\vec \varphi )\right] - 
\nonumber \\
&&\ -\frac{RL}{N_\sigma ^3}\sum_{\vec \varphi }\frac{\beta _1^{\prime }\cos
\varphi _1}{s_0-A(\vec \varphi )},  \label{part2}
\end{eqnarray}
therefore from 
\begin{equation}
W_{R\times L}=\exp \left\{ -RLa_2a_3\sigma ^{(1)}\right\} ,  \label{W-sm}
\end{equation}
we have for the string tension 
\begin{equation}
\sigma ^{(1)}a_2a_3=\beta _1^{\prime }\int \left( \frac{d\varphi }{2\pi }%
\right) ^3\frac{\cos \varphi _1}{s_0-\sum_{\vec n}\beta _n^{\prime }\cos
\varphi _n},  \label{string}
\end{equation}
This equation can be rewritten as 
\begin{equation}
\sigma ^{(1)}a_2a_3=\frac{\beta _1^{\prime }}{N_\sigma ^3}\frac \partial {%
\partial \beta _1^{\prime }}\ln Z_0.  \label{dwriv}
\end{equation}

Since the string tension in a certain sense characterizes the average
plaquette value , one may anticipate that this quantity is defined (through
the duality relations) in the terms of the average value of the link ${\frak %
L}\equiv \frac 1{N_\sigma ^3}\frac \partial {\partial \beta _1^{\prime }}\ln
Z_0$

For the 'large' Wilson loop $(R\approx N_2;L\approx N_3)$ 
\begin{equation}
\Delta (\varphi _n;\varphi _m^{\prime })\approx 2RL\delta (\varphi
_2-\varphi _2^{\prime })\delta (\varphi _3-\varphi _3^{\prime }),
\label{large}
\end{equation}
we have 
\begin{eqnarray}
\  &&N_\sigma ^3\left[ \delta _{\vec \varphi ^{\prime }}^{\vec \varphi }A(%
\vec \varphi )-\frac{\beta _1^{\prime }}{N_\sigma ^3}\Delta (\varphi
_n;\varphi _m^{\prime })\cos \varphi _{^{\prime }1}\right] =  \nonumber \\
&&\ N_\sigma ^3\delta _{\varphi _2^{\prime }}^{\varphi _2}\delta _{\varphi
_3^{\prime }}^{\varphi _3}\left[ \delta _{\varphi _1^{\prime }}^{\varphi
_1}A(\vec \varphi )-\frac{\beta _1^{\prime }RL}{N_\sigma ^3}2\cos \varphi
_1^{\prime }\right] ,
\end{eqnarray}
and the free energy $F\equiv -\frac{\ln Z}{N_\sigma ^3}$ can be written 
\begin{eqnarray}
\  &&-F=\int \left( \frac{d\varphi }{2\pi }\right) ^3\ln A(\vec \varphi )+ 
\nonumber \\
&&\ +\frac 1{N_1}\int \frac{d\varphi _2d\varphi _3}{\left( 2\pi \right) ^2}%
\ln \left[ 1-\frac{\beta _1^{\prime }RL}{N_2N_3}\int \frac{d\varphi _1}{2\pi 
}\frac{\cos \varphi _1}{A(\vec \varphi )}\right] .  \label{large2}
\end{eqnarray}
The string tension is expressed as 
\begin{equation}
\sigma ^{\left( 1\right) }a_2a_3=\int \frac{d\varphi _2d\varphi _3}{\left(
2\pi \right) ^2}\ln \left[ 1-\frac{\beta _1^{\prime }RL}{N_2N_3}\int \frac{%
d\varphi _1}{2\pi }\frac{\cos \varphi _1}{A\left( {\bf \varphi }\right) }%
\right] .  \label{string2}
\end{equation}
It is easy to see that for $\frac{RL}{N_2N_3}<<1$ we come back to $(\ref
{string})$

After the integration over $\varphi _1$ and $\varphi _2$, one can easily get 
\begin{equation}
\sigma ^{(1)}a_2a_3=\zeta (s_0)-\frac{\zeta (s_0-\beta _1^{\prime })+\zeta
(s_0+\beta _1^{\prime })}2,
\end{equation}
where 
\begin{eqnarray}
&&\zeta (s_0)=\int_0^\pi \frac{d\varphi _3}\pi  \\
&&\ln \left( s_0-\beta _3^{\prime }\cos \varphi _3+\sqrt{(s_0-\beta
_3^{\prime }\cos \varphi _3)^2-(\beta _2^{\prime })^2}\right) .
\end{eqnarray}

In the particular case $(\beta _n^{\prime }=\beta ^{\prime };~~~a_n=a)$ 
\begin{equation}
\sigma ^{\left( 1\right) }a^2\equiv \sigma a^2=\zeta \left( s_0\right) -%
\frac 12\zeta \left( s_0-\beta ^{\prime }\right) -\frac 12\zeta \left(
s_0+\beta ^{\prime }\right) ,  \label{str}
\end{equation}
and we obtain 
\begin{equation}
\sigma a^2\approx \left\{ 
\begin{array}{ll}
c_0+O\left( \beta ^{\prime }-\beta _c{}^{\prime }\right) ^2; & s_0\gtrsim
3\beta _c^{\prime }\approx \frac 32, \\ 
\frac 12(\beta ^{\prime })^2+\frac{3(\beta ^{\prime })^4}8; & \frac{\beta
^{\prime }}{s_0}<<1;
\end{array}
\right. 
\end{equation}
where the constant $c_0$ insignificantly differs for the 'small' $\left(
c_0\approx \frac 13\right) $ and for the 'large' $\left( c_0\approx \frac 14%
\right) $ loops.

In the more general case $a_2=a_3\neq a_1$ 
\begin{equation}
\sigma ^{\left( 1\right) }\approx \sigma \cdot \left( \frac{\beta _1^{\prime
}}{\beta ^{\prime }}\right) ^2\left( 1+O(\beta _n^{\prime })^2\right) .
\label{st}
\end{equation}

It is easy see that in order to obtain the corresponding expression for $%
Z\left( 3\right) $ the gauge group in given approximation one should change
only coupling constant in $\left( \ref{st}\right) $ given by $\left( \ref
{c-ng}\right) .$

As one may anticipate for 'extremely small' loops $\left( R\sim 1;L\sim
1\right) $ e.g. single plaquette, approximate expressions for 'small' loops $%
\left( \ref{W-sm}\right) $and $\left( \ref{deriv}\right) $ become exact.

In a spherical model approximation 'Creutz ratios' are the same for 'large'
and 'small' Wilson loops 
\begin{eqnarray}
&&-\ln \frac{W\left( R,L\right) W\left( R-1,L-1\right) }{W\left(
R,L-1\right) W\left( R-1,L\right) }=  \nonumber \\
&=&\sigma ^{(1)}a_2a_3=\beta _1^{\prime }\int \left( \frac{d\varphi }{2\pi }%
\right) ^3\frac{\cos \varphi _1}{A(\vec \varphi )}.
\end{eqnarray}
It should be noted, however, that the difference between 'large' and 'small'
Wilson loops has nothing to do with finite size effects, it will survive
even at an infinite lattice and reflect alternity between finite Wilson
loops and infinite ones. At least in given approximation, such difference is
not so dramatic, as it was forewarned in \cite{kowall}.

Comparison with the MC experiment show that spherical model predictions {\it %
qualitively} agree with it . Nonanalitical, but quite smooth behavior is
demonstrated near the critical point $\sigma \left( \beta ^{\prime }\right) $%
. In the critical region $\beta ^{\prime }\sim \beta _c^{\prime }$ the value 
$\sigma \left( \beta ^{\prime }\right) \approx $ $\sigma ^c$. In the deep
deconfinement region $\beta ^{\prime }<<\beta _c^{\prime }$ (where saddle
point steadily moved $s_0\rightarrow 1$ ) $\sigma \left( \beta ^{\prime
}\right) $ decrease with $g^2\rightarrow 0$ smoothly, but too fast if
compared with the MC experiment.

\subsection{Spatial string tension}

Here we consider the case of $a_n=a;$ $a_0=a_\tau $ $=a/\xi $ and put 
\begin{equation}
\beta =\frac 2{g^2}\frac 1\xi ;\quad \beta _\tau =\frac 2{g^2}\xi =\beta \xi
^2,
\end{equation}
therefore the expression $\left( \ref{st}\right) $ for string tension in
spherical model approximation can be rewritten as

\begin{equation}
\sqrt{\sigma }\approx \sqrt{\frac 12}\cdot \frac{\beta ^{\prime }}a
\label{STR}
\end{equation}
which in strong coupling region with $\left( \ref{c-ng}\right) $or $\left( 
\ref{c-ng-p}\right) $ leads to 
\begin{equation}
\sqrt{\sigma }\approx const\cdot \frac{\ln \beta }a,  \label{STsc}
\end{equation}
therefore $\sqrt{\sigma }$ will scale if we demand $\beta \simeq \exp \left(
a\times const\right) $

In weak coupling region the expression $\left( \ref{STR}\right) $ with $%
\left( \ref{c-ng}\right) $ gives 
\begin{equation}
\sqrt{\sigma }\approx \left\{ 
\begin{array}{cc}
\frac{a^{-1}}{\sqrt{2}}e^{-2\beta }; & Z\left( 2\right) , \\ 
\frac{a^{-1}}{\sqrt{2}}e^{-\frac 32\beta }; & Z\left( 3\right) 
\end{array}
\right.   \label{st2}
\end{equation}
or 
\begin{equation}
\sqrt{\sigma }\approx \frac{a^{-1}}{\sqrt{2}}\exp \left\{ -\beta \left(
1-\cos \frac{2\pi }N\right) \right\}   \label{stN}
\end{equation}
and taking into account 
\begin{equation}
\quad g_\sigma ^{-2}\approx b_0\ln \frac 1{a_\tau \Lambda _L}+...  \label{G}
\end{equation}
obtained in \cite{hasen1} and \cite{shig} we finally get

\begin{equation}
\sqrt{\sigma }\backsim a_\tau ^\epsilon  \label{sigma}
\end{equation}
with 
\begin{equation}
\epsilon \equiv \frac{2N}\xi \cdot \left( 1-\cos \frac{2\pi }N\right) b_0-1.
\label{sigmaN}
\end{equation}
For $\xi =1$ the scaling condition $\epsilon =0$ leads to 
\begin{equation}
b_0=\frac{1/2N}{\left( 1-\cos \frac{2\pi }N\right) }\approx \frac N{4\pi ^2}%
\approx .025N\ ,
\end{equation}
which for large enough $N$ agrees with the standard value $b_0=\frac{11}{12}%
\frac N{4\pi ^2}\approx .023N$, so all this looks as if the spatial string
tension on asymmetric lattice ($\xi <<1$) acquires 'anomalous dimension' $%
\epsilon $ , which disappears when \ $\xi \rightarrow 1.$

\section{Discussion}

The application of spherical model in statistical physics has long history
since the time it has been introduced to investigate critical phenomena in
the ferromagnet \cite{kac} and until now ( see, e.g. \cite{Fra-Hen}, \cite
{Cap-Col}). Although this model is of no direct experimental relevance, it
may provide useful insight since many physical quantities of interest can be
exactly evaluated with its help. In this context, the spherical model is
quite a useful tool in providing explicit verification of general concepts
in critical phenomena, see \cite{Barb}\cite{Sing}\cite{Coni}\cite{Khor}.

\ Recently \cite{Fra-Hen} it was successfully used for studying the
transitions between a paramagnetic, a ferromagnetic and an ordered
incommensurate phase (Lifshitz point). Spherical model approximation helped
to find the exact scaling function of a system with strongly anisotropic
scaling. Models of this kind were investigated extensively (see recent
review in \cite{revselke}).

As it has been established by Stanley \cite{Stan}, there is precise
correspondence between the spherical model and the Heisenberg model. Indeed,
consider a $d$ - dimensional lattice of $N$\ classical spin 
\begin{eqnarray}
{\rm Z}_N^{\left( \nu \right) } &=&\int_{-\infty }^{+\infty }\prod_l\delta
\left( \nu -\left| U_l^{\left( \nu \right) }\right| ^2\right) \times 
\nonumber \\
&&\exp \left( -\sum_{l,l^{\prime }}J_{ll^{\prime }}U_l^{\left( \nu \right)
}U_{l^{\prime }}^{\left( \nu \right) }\right) \prod_ldU_l^{\left( \nu
\right) }.
\end{eqnarray}
where $U_l^{\left( \nu \right) }$\ is a $\nu $\ - dimensional vector of
length $\nu ^{\frac 12}$, and ${\rm Z}_N^{\left( \nu \right) }$ is
corresponding partition function. In the limit $N,\nu \rightarrow \infty $
the free energy of this classical Heisenberg model is identical to that of
the spherical model \cite{Stan}. Kac and Thompson clarified the situation by
proving rigorously that this model gives a surprisingly good result for any
fixed temperature above and below critical point and is independent of the
ordering of the limits $N\rightarrow \infty $\ and $\nu \rightarrow \infty .$

In \cite{Cap-Col} it was shown that the spherical model is associated with
the matrix model \cite{matmod}, which has the same diagrammatic expansion
and saddle points in the planar approximation $XY$\ model. The analogy with
matrix models is interesting because it could provide some useful technology 
\cite{matmod} for solving the $XY$\ model. In two recent papers \cite
{parisi1}, \cite{parisi2}, Parisi et al. introduced and analyzed the
spherical and $XY$\ spin models with frustration to test the conjecture that
the frustrated deterministic systems at low temperature behave similar to
some suitably chosen spin-glass models with quenched disorder \cite
{spinglass}

Spherical model predicts reasonable values for critical exponents \cite
{joyce} Moreover, a 'basic' set of exponent relations is also satisfied by
spherical model for $d<5$.

An advantage of the spherical model is that it satisfactorily describes
fluctuations and therefore is suitable for the computation of correlation
functions either for fixed lattice volume or for infinite one. As it was
pointed out by \cite{seiler} on lattices with infinite volume the
perturbation theory (which is in fact a saddle point expansion around an
ordered state) gives ambiguous results, and certainly fails for some border
conditions.

\section{Conclusion}

Spatial string tension for 'small' and 'large' loops, as one can expect,
shows only marginal difference, which in Creutz ratio can not be discerned
at all. This difference, however, is not a lattice artefact and, in
principle, can be detected in MC experiment. Spatial string tension power
dependence on lattice spacing, demanded at $\left( \ref{sigma}\right) ,$
formally does not contradict to condition obtained by standard
renormalisation procedure, however, the explicit form of the dependence $%
\sqrt{\sigma }$\ on $\xi $\ at $\left( \ref{sigma}\right) $\ disagrees with
it.\vspace{1.0in}%
%
%
%
%

\end{document}